\documentclass[twocolumn,showpacs,preprintnumbers,amsmath,amssymb]{revtex4}

\usepackage{graphicx}
\usepackage{dcolumn}
\usepackage{bm}

\begin{document}

\title{Magnetic remanent states in antiferromagnetically coupled multilayers
}

\author{N. S. Kiselev$^{a,b}$}
\thanks
{Corresponding author 
}
\email{m.kyselov@ifw-dresden.de}

\author{U. K. R\"o\ss ler$^a$}
\author{A. N. Bogdanov$^a$}
\author{O. Hellwig$^c$}
\affiliation{$^a$IFW Dresden, Postfach 270116, D-01171 Dresden, Germany}
\affiliation{$^b$Donetsk Institute for Physics and Technology, 
 83114 Donetsk, Ukraine}
\affiliation{$^c$San Jose Research Center, Hitachi Global Storage Technologies,
 San Jose, CA 95135, USA}


\begin{abstract}
In antiferromagnetically coupled multilayers with
perpendicular anisotropy 
unusual multidomain textures can be stabilized
due to a close competition between 
long-range demagnetization fields 
and short-range interlayer exchange coupling.
In particular, the formation and evolution
of specific topologically stable planar defects
within the antiferromagnetic ground state, i.e. 
wall-like structures with a ferromagnetic configuration 
extended over a finite width,
explain configurational hysteresis phenomena
recently observed in [Co/Pt(Pd)]/Ru and [Co/Pt]/NiO multilayers.
Within a phenomenological theory, 
we  have analytically derived the equilibrium sizes of 
these ``ferroband'' defects as functions 
of the antiferromagnetic exchange, a bias magnetic field,
and geometrical parameters of the multilayers.
In the magnetic phase diagram,  
the existence region of the ferrobands mediates 
between the regions of patterns
with sharp antiferromagnetic domain walls 
and regular arrays of ferromagnetic stripes.
The theoretical results are supported by 
magnetic force microscopy images
of the remanent states observed in [Co/Pt]/Ru.
\end{abstract}

\maketitle
\section{Introduction}
Recently synthesized antiferromagnetically coupled 
multilayers with strong perpendicular magnetic anisotropy,
such as [Co/Pt]/Ru, [Co/Pt]/NiO, Co/Ir, Fe/Au,
behave as \textit{synthetic metamagnets}.
Such perpendicular media
are currently investigated 
as  promising materials for thermally 
stable high-density recording technologies 
and for the emerging spin electronics.
These effectively antiferromagnetic structures
display a rich variety of magnetization reversal processes 
accompanied by various reorientation effects 
and the formation of complex multidomain structures \cite{Hellwig07}.
In addition to the equilibrium multidomain phases
different systems of irregular networks of domain
walls and bands have been reported to exist in 
the antiferromagnetic phase of [Co/Pt]/Ru \cite{Hellwig07,%
Hellwig03a,Hauet08}, [Co/Pt]/NiO \cite{Baruth06,Liu08} 
and [Co/Pd]/Ru \cite{Fu07} multilayers.
These topologically stable planar magnetic 
defects strongly depend on the magnetic 
and temperature history
\cite{Hellwig07,Hauet08,Fu07,Davies08}.
Micromagnetic analysis shows that,
depending on the material parameters and the 
magnetic history, the antiferromagnetic remanent 
states may display sharp domain walls, 
``trapped'' ferromagnetic strips which we call \textit{ferrobands},
antiferromagnetic strips in a metastable ferrostripe matrix, 
and various metastable isolated defects 
formed from remanent domains with internally ferrimagnetic state 
\cite{topdefects08}.
The formation of these magnetic defects
leads to configurational hysteresis, 
as has been observed in 
the remanent states of antiferromagnetically coupled multilayers 
\cite{Hellwig07,Hauet08,Fu07}.
In this paper we apply a phenomenological 
theory developed in \cite{topdefects08} 
to determine the existence regions in the 
magnetic phase diagram and the equilibrium 
sizes for the ferroband defects.
The theoretical results explain 
recent experimental observations
of remanent defect states
in various perpendicular antiferromagnetic 
multilayer systems \cite{Hellwig07,%
Hellwig03a,Hauet08,Baruth06,Liu08,Fu07,Davies08}.
The theoretical approach creates 
a consistent and quantitative micromagnetic 
model for the domain patterns responsible
for configurational hysteresis effects 
in this class of materials.
%

\section{Ferrobands versus sharp domain walls}

\begin{figure}
\includegraphics[width=8.5 cm]{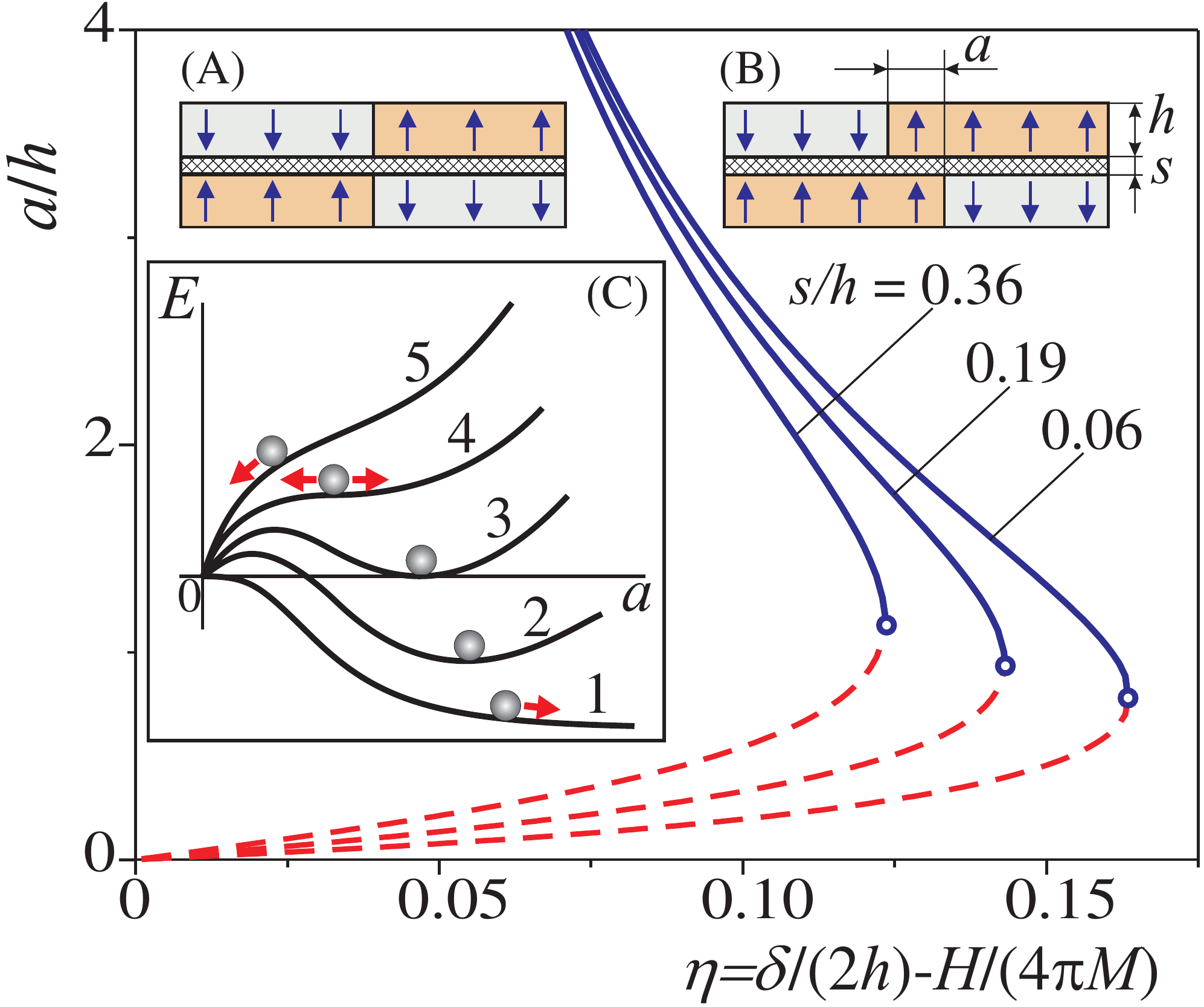}
\caption{
\label{f_1}
(Color online)
The optimal values of the reduced width
$a/h$ of ferrobands (Inset B) as functions
of the effective magnetic coupling parameter $\eta$,
combining strength of interlayer exchange $\delta$ and
bias field $H$,
in a bilayer $N=2$ (solid lines). 
Dashed lines indicate unstable solutions.
Hollow points mark critical values of
ferrobands ($\eta_{cr}, a_{cr}$).
Below $a_{cr}(s/h)$ only sharp domain
walls (Inset A) exist.
Inset C shows the systematic deformation 
of the profiles for the energy vs. 
ferroband width $a$ under a decreasing bias field from 1 to 5,
see text for details.
}
\end{figure}

According to the 
experimental observations 
and theoretical analysis,
the antiferromagnetic multilayers with
strong perpendicular anisotropy may have
antiferromagnetic single-domain structure 
as zero-field ground-state in certain ranges of 
geometry and materials parameters
\cite{Hellwig07,APL07}.
Planar defects separating antiferromagnetic domains 
in remanent states of these multilayers
may arise either 
as \textit{sharp} domain walls 
or as \textit{ferrobands} 
(``shifted antiferromagnetic walls'') (Fig. \ref{f_1} Insets (A) and (B)).
%
%
Sharp walls are similar to 180$^\circ$ 
domain walls in bulk antiferromagnets.
Ferrobands arise in antiferromagnetically 
coupled multilayers due to a subtle 
interplay between magnetodipole 
and interlayer exchange interactions.
To investigate this phenomenon  we consider 
an isolated ferroband of width $a$ in a
multilayer consisting of $N$ identical magnetic layers
of thickness $h$ separated 
by nonmagnetic spacers of thickness $s$ (Fig.~\ref{f_1}, Inset (B)).
The magnetic energy of this system (per unity band length)
can be written in the following form \cite{APL07}
\begin{eqnarray}
E = 4 \pi M^2 h^2 N
\left[ F (u) + \eta u  \right] \,,
\label{energy0}
\end{eqnarray}
where $u= a/h$ is the reduced band width,
the  magnetostatic energy $F(u)$ is derived 
by solving the corresponding
magnetostatic problem for a ``charged'' band 
\begin{eqnarray}
\label{energyband}
F(u) &=& \frac{1}{4 \pi} \sum\limits_{\scriptstyle k = 1 \hfill \atop 
  \scriptstyle odd \hfill}^{N - 1} {\left( {1 - \frac{k}{N}} \right)} 
  \Xi \left(u, \tau k\right)
\end{eqnarray}
where $\tau=1+s/h$, and
\begin{eqnarray}
\Xi \left(u, \tau k\right) &=& 2 f (u, \tau k) - \!f(u, \tau k+\!1) - \!f (u,\tau k-\!1),\\
\nonumber 
f(u,\omega)   &=&(\omega  ^2-u^2)\ln  ( \omega^2 + u^2)-\\
\nonumber
\, & &\omega^2\; \ln(\omega^2) - 4 \omega u \arctan (u/\omega)\,.
\end{eqnarray}
Here we introduce an effective magnetic coupling parameter
\begin{eqnarray}
\eta = \left(1- \frac{1}{N} \right)\frac{\delta}{h} 
-\frac{H}{4 \pi M}\,.
\label{eta}
\end{eqnarray}
$H$  is an applied magnetic field 
perpendicular to the multilayer.
The \textit{exchange length} $\delta$
is given by the ratio of 
the antiferromagnetic coupling $J>0$ and the stray-field energy,
$\delta = J/(2\pi M^2)$.
Note that $\eta$ includes all material parameters
of the systems, while the reduced magnetostatic energy
$F(u)$ depends only on geometrical parameters of
the multilayer, namely, the ratio $s/h$.
The condition $d E / d u = 0$ yields the equation 
for equilibrium ferroband widths:
\begin{eqnarray}
\eta = G(u) \equiv
\frac{1}{8 \pi} && \sum \limits_{\scriptstyle k = 1 \hfill \atop 
  \scriptstyle odd \hfill}^{N - 1} {\left( \!{1\! -\! \frac{k}{N}} \!\right)} 
  \Xi_u \left(u, \tau k\right)\,,
\label{equationband}
\end{eqnarray}
where
\begin{eqnarray}
\Xi_u & =&2 g (u, \tau k) - g(u, \tau k+1) - g (u,\tau k-1),
\\
\nonumber 
g(u,\omega) & =&  2\left[ {u\ln \left( {\omega ^2  + u^2 } \right) + u + 
2\omega \arctan \left( {\frac{u}{\omega }} \right)} \right].
\end{eqnarray}

Typical solutions of Eq. (\ref{equationband}) are plotted in
Fig.~\ref{f_1} for thickness ratios $s/h$ corresponding
to geometrical parameters in different experimentally 
investigated systems:
$s/h$ = 0.36 \cite{Fu07}, 0.19 \cite{Hellwig07} and 0.06 \cite{Liu08}.
Note that a sharp domain wall can be treated as the limiting 
case of a ferroband with zero width.
%
\begin{figure}
\includegraphics[width=8.5 cm]{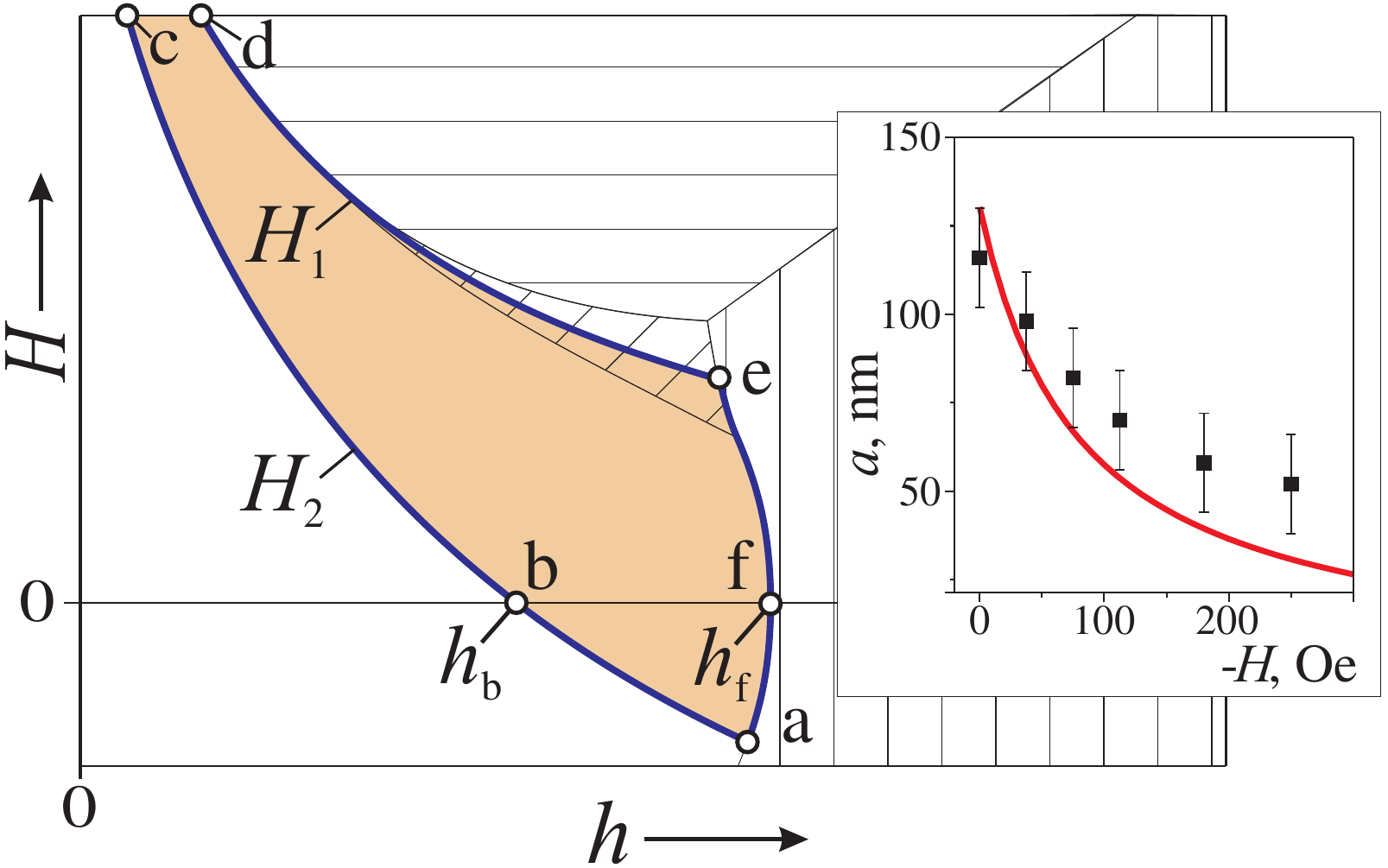}
\caption{
\label{f_2}
(Color online) The magnetic phase diagram of a bilayer (schematically)
displaying the existence region of metastable ferroband
defects (filled area) and the equilibrium states:
(i) the homogeneous antiferromagnetic state (white area), 
(ii) the ferromagnetic state (horizontally hatched),
(iii) the multidomain ferrostripe state  (vertical hatched),
and (iv) the  metamagnetic state (slanting hatched area).
At the critical line $H_1$ ($d-e$) the ferroband transform 
into saturated states by an unlimited expansion of their sizes.
At the line  $H_2$ ($a-b-c$) 
the ferrobands transform into sharp domain walls.
The $a-f-e$ line marks the transition into
the ferrostripe phase.
The inset shows the ferroband width as a function
of an opposing (negative) magnetic field (solid line)
for a [Co/Pt]NiO bilayer investigated in Ref.~\cite{Liu08}.
(Line for present theory, points experimental data \cite{Liu08}). 
}
\end{figure}

The stability condition, $d^2E/du^2=0$, gives the equation
\begin{eqnarray}
\label{eqtwo}
\sum\limits_{\scriptstyle k = 1 \hfill \atop 
  \scriptstyle odd \hfill}^{N - 1} \!\! {\left(\! {1 - \frac{k}{N}} \!\right)} 
   \ln\!\! 
   \left[ 1 + \frac{1 + 2 u^2 - 2 \tau^2 k^2}{(\tau^2 k^2 + u^2)^2}  \!\right]\!=\!0\,
\end{eqnarray}
which, combined with Eq.~(\ref{equationband}),
determines critical values of the 
ferroband width $u_{c}$ and $\eta_{c}$.
Because Eq.~(\ref{eqtwo}) does not 
include the material parameters, the solutions
for $u_{c}$ are functions of the ratio $s/h$ alone.
In particular, for  bilayers 
 \begin{eqnarray}
\label{ac}
a_{c} =\sqrt{s^2 + 2 s h +h^2/2}.
\end{eqnarray}
By substituting $u_{c}$ into Eq.~(\ref{equationband})
we find $\eta_{c} = G_c \equiv G (u_c) $.
The analysis shows that solutions of Eq.~(\ref{equationband})
exist in the range $ 0 < \eta < \eta_{c}(s/h)$.
The equations $\eta =0$ and $\eta = \eta_{c}(s/h)$
yield the limiting fields where ferroband 
solutions can exist as stable defects (Fig.~\ref{f_2})
 \begin{eqnarray}
\label{H1}
 & & H_1 (h) = 4\pi M  \left( 1 - \frac{1}{N} \right) \frac{\delta}{h},
\nonumber \\
 & & H_2 (h_c)= H_1(h_c)- G_c (s/h_c).
\end{eqnarray}
Inset (C) in Fig.~\ref{f_1} shows the evolution of the energy profiles
$E(a)$ when the bias field varies from $H = H_1$ (profile 1) to
$H = H_2$ (profile 4). At the critical field $H_1$ (line $d-e$ in Fig.~\ref{f_2} )
the energy of the antiferromagnetic phase equals the energy 
of the ferromagnetic (saturated) state. 
Here, both  sharp domain walls ($a =0$) solutions 
and ferrobands expand to infinity.
For lower fields both topological defects are locally
stable (profile 2). 
In decreasing bias field the ferroband energy gradually increases. 
At a certain value of the bias
field $H^* (h)$ the energies of both defect types 
become equal (profile 3), 
and for  $H^* (h) > H > H_2(h)$ the ferroband
energy is larger than that of the sharp domain wall.
Finally at the critical field $H_2(h)$ the ferrobands collapse (profile 4), 
and for $H <H_2(h)$ only sharp wall solutions
can exist (profile 5).

Because the variation of the ferroband width 
does not change the domain wall energy
the equilibrium ferroband sizes
do not depend on the characteristic length.
They are formed only under competing influence
of the antiferromagnetic exchange and the combined
external bias and dipolar stray fields. 

\section{Reorientation effects and remanent states}

Topological defects 
can not arise spontaneously. 
However, they can be induced by magnetization processes.
Thus, the formation and evolution of topological defects
strongly depends 
on the sequence of magnetic-field-driven states 
and the transitions between them.
Usually antiferromagnetic domain walls
and ferrobands  arise in the remanent state
after demagnetization \cite{Hellwig07,Fu07}.
This follows from the fact that
antiferromagnetic domain walls and ferrobands
are remnants of the ferromagnetic
phases within the antiferromagnetic matrix.
These wall defects also arise after in-plane demagnetization
however, the defected antiferromagnetic state, i.e. the domain pattern 
may own different sizes and morphologies after different field histories \cite{Hellwig07}.
But, the structure of 
the topological wall defects should be the same in either case.
Depending on the magnetic layer thickness remanent
states consist of multidomain patterns with sharp domain
walls ($h <h_b$), ferrobands ($h_b < h < h_f$),
or the regular ferrostripe phase ($h > h_f$)
(Fig. 3).
In multilayers with thicker magnetic layers
the antiferromagnetic and ferromagnetic phases
are separated by the region of transitional
domain structures (\textit{metamagnetic} domains)
(Fig.~\ref{f_2}).  
In this case the antiferromagnetic phase may 
include remnants of metamagnetic domains.
Such textures have been observed in [Co/Pt]Ru
multilayers after out-of-plane saturation
\cite{Hellwig07}.
Experimental data on the variation of the ferroband size 
under influence of the applied field
have been reported in Ref.~\cite{Liu08}. 
In the experimental investigation 
a ferroband in a 
[Pt(5\AA)/Co(4\AA)]$_4$/NiO(11\AA)/[Co(4\AA)/Pt(5\AA)]$_4$ 
bilayer was squeezed by an opposing magnetic field.
By fitting the experimental data of Ref.~\cite{Liu08}
we calculate from Eq.~(\ref{equationband})
the exchange constant 
$J =0.002$ erg/cm$^2$ ($\delta = 0.063$ nm) 
and the optimal ferroband
width as a function of the bias field, $a(H)$ (Fig.~\ref{f_2}, Inset).
This value of  $J$ is in reasonable agreement with those for
[Co/Pt]NiO multilayers investigated in
Ref.  \cite{Baruth06}.

According to experimental observations 
ferrobands can exist either in 
a single domain state \cite{Baruth06,Liu08,Fu07}
or split into a system of domains creating, 
so called ``tiger-tail'' patterns \cite{Hellwig07,Hellwig03a}.
The ``tiger-tail''  multidomain states of these defects
clearly are due to dipolar depolarization. 
In principle, these effects can be considered by additional
stray field terms for the modulated magnetization 
along a ferroband in the model energy (\ref{energy0}).
Further experimental investigations of ``tiger-tail'' patterns
together with a micromagnetic analysis of these multidomain
patterns should give deeper insight into the formation 
and evolution of topological defects in this class of
magnetic nanostructures.
\begin{figure}
\includegraphics[width=8.5 cm]{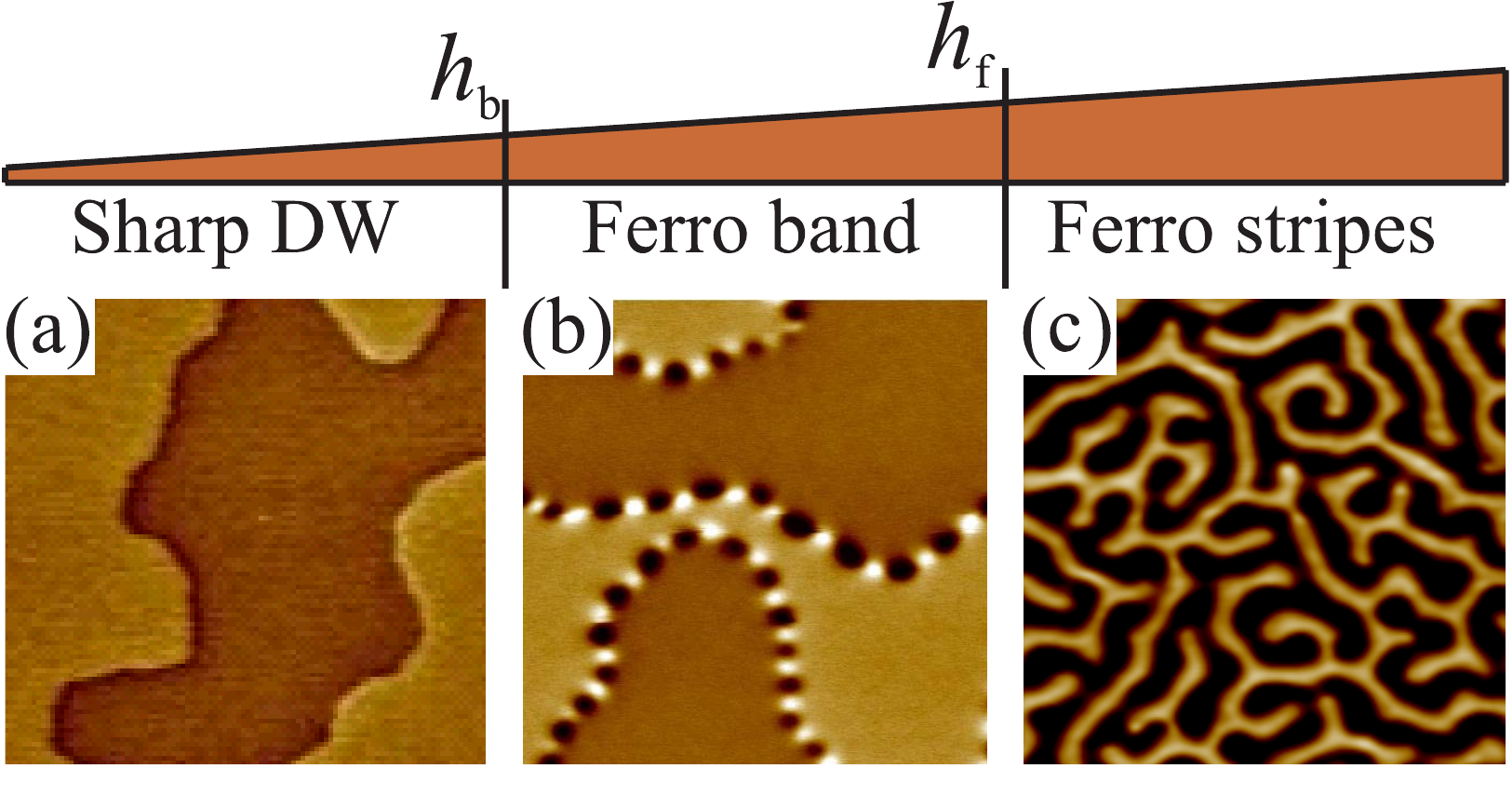}
\caption{
\label{f_3}
(Color online)
Remanent states observed in 
Pt(20)[[Co(h)/Pt(0.7)]$_7$ Co(h) Ru(0.9)]$_{14}$
[Co(h)Pt(0.7)]$_8$ Pt(0.13) multilayers
with wedged Co layers after
in-plane demagnetization
(a, b) and out-of-plane saturation (c) 
($h$ = 0.36 - 0.4 nm for a, $h$ =
0.4 - 0.44 nm for b and c),
field of view is 3.5 $\mu$m $\times$ 3.5 $\mu$m).
For thin layers $h < h_b$ 
antiferromagnetic patterns with sharp domain
walls (a) occur;
for $h_b < h < h_f$ antiferromagnetic domains
are separated by ferrobands (b) splitted to
up and down domains 
(so called, "tiger-tail" patterns).
For thicker magnetic layers
($h > h_b$) the remanent state consists of
thermodynamically stable ferrostripes.
}
\end{figure}
%
%

In conclusion, we present an exhaustive analysis of
specific topological defects (ferrobands) arising
in perpendicular antiferromagnetically coupled multilayers.
Our analytical solutions generalize and complete numerical
studies of these defects in Refs. \cite{Hellwig07,Hellwig03a,Hauet08,%
Baruth06}.
Magnetic-field-driven evolution and transformation of ferrobands
explain the formation of defected remanent states
recently observed in [Co/Pt]Ru 
and [Co/Pt]NiO antiferromagnetic multilayers.

\section{Acknowledgment} Work supported by DFG SPP1239 project A8.
N.S.K. and A.N.B. thank H. Eschrig for the support and the hospitality
at IFW Dresden.

\maketitle

\end{document}